\documentclass[galaxies,article,submit,moreauthors,10pt,letterpaper]{mdpi_no_lineno}
\pdfoutput=1
\firstpage{1}
\articlenumber{x}
\doinum{10.3390/------}
\pubvolume{xx}
\pubyear{2016}
\copyrightyear{2016}
\externaleditor{Academic Editors: Jose L. G\'{o}mez, Alan P. Marscher and Svetlana G. Jorstad}
\history{Received: 1 September 2016; Accepted: 8 October 2016; Published: date}

\theoremstyle{mdpi}
\newcounter{thm}
\newcounter{ex}
\newcounter{re}
\usepackage{soul}
\usepackage{microtype}
\newcommand\myurl[1]{\changeurlcolor{black}\url{#1}\changeurlcolor{blue}}
\makeatletter
\g@addto@macro{\UrlBreaks}{\UrlOrds}
\makeatother

\newcommand {\msun}{M$_{\odot}$}
\newcommand{\degree}{\ensuremath{^\circ}}

\Title{Observations of the Structure and Dynamics of the Inner M87 Jet}

\Author{R. Craig Walker $^{1,}$*, Philip E. Hardee $^{2}$, Fred Davies
$^{3}$, Chun Ly $^{4}$, William Junor ${^5}$, \mbox{Florent~Mertens $^{6,7}$} and
Andrei Lobanov $^{6,8}$}

\AuthorNames{R. Craig Walker, Philip E. Hardee, Fred Davies, Chun Ly,
William Junor, Florent Mertens, and Andrei Lobanov}

\address{%
$^{1}$ \quad National Radio Astronomy Observatory, Socorro, NM 87801, USA\\
$^{2}$ \quad Department of Physics \& Astronomy, The University of Alabama,
Tuscaloosa, AL 35487, USA; pehardee@gmail.com\\
$^{3}$ \quad MPIA, 69117 Heidelberg, Germany; fdavies@ucla.edu\\
$^{4}$ \quad Goddard Space Flight Center, Astrophysics Science Division, 8800
Greenbelt Road, Greenbelt, \mbox{MD 20771, USA;} astro.chun@gmail.com\\
$^{5}$ \quad ISR-2, MS-D436, Los Alamos National Laboratory,
P.O. Box 1663, Los Alamos, NM 87545, USA; bjunor@lanl.gov\\
$^{6}$ \quad Max Plank Institute f{\"u}r Radioastronomie, Auf dem Huegel 69,
53121 Bonn, Germany; alobanov@mpifr-bonn.mpg.de\\
$^{7}$ \quad Kapteyn Astronomical Institute, PO Box 800, 9700 AV
Groningen, The Netherlands; florent.mertens@gmail.com\\
$^{8}$ \quad Institut f{\"u}r Experimentalphysik, Universit{\"a}t Hamburg,
Luruper Chaussee 149, 22761 Hamburg, Germany}

\corres{Correspondence: cwalker@nrao.edu}

\abstract{M87 is the best source in which to study a jet at high
resolution in gravitational units because it has a very high mass
black hole and is nearby.  The angular size of the black hole is
second only to Sgr~A*, which does not have a strong jet.  The jet
structure is edge brightened with a wide opening angle base and a weak
counterjet.  We have roughly annual observations for 17 years plus
intensive monitoring at three week intervals for a year and five day
intervals for 2.5~months made with the Very
Long Baseline Array (VLBA) at 43 GHz.  The inner jet
shows very complex dynamics, with apparent motions both along and
across the jet.  Speeds from zero to over 2c are seen, with
acceleration observed over the first 3 milli-arcseconds.  The
counterjet decreases in brightness much more rapidly than the main
jet, as is expected from relativistic beaming in an accelerating jet
oriented near the line-of-sight.  \mbox{Details of the} structure and
dynamics are discussed.  The roughly annual observations show
side-to-side motion of the whole jet with a characteristic time scale
of about 9~years.}


\begin{document}

\section{Introduction}

M87 is a dominant galaxy of the Virgo Cluster at a distance of 16.7
Mpc \cite{Mei07}.  There is a jet emitted from the nucleus that is
visible throughout the electromagnetic spectrum.  It is a strong radio
source, known as Virgo A, with jet and lobe structures spanning the
smallest scales probed by high frequency Very Long Baseline
Interferometry (VLBI) to arcminute scales probed by instruments such
as the Jansky Very Large Array and LOFAR.
  M87 contains a very massive
black hole.  The~mass is still uncertain with values of M$_{BH} =
(3.5^{+0.9}_{-0.7}) \times 10^9$~\msun~ determined from gas dynamics
\cite{Walsh2013} and $(6.6 \pm 0.4) \times 10^9$~\msun~ determined
from stellar dynamics \cite{G2011}.  In this work, we use a mass of
$6.16 \times 10^9~$\msun~which is the stellar dynamic mass adjusted
for our assumed distance.  The assumed mass and distance give a
Schwarzschild radius (R$_s \equiv 2GM/c^2$) of 122 au or 7.3
${\mu}$arcsec.  This is the highest angular size black hole that has a
radio jet whose structure can be studied and is the second highest
angular size black hole associated with any radio source, second only
to Sgr~A* in the center of the Milky Way.  Thus M87 is the best source
in which to observe the jet base region where the jet is accelerated
and collimated.

The M87 jet was reported to have a wide-opening-angle base and edge
brightened structure all the way to the core as seen by 43 GHz Very
Long Baseline Array (VLBA \cite{Napier1993}) observations
\cite{J1999}.  A~counterjet has been seen in most VLBA 43 GHz images
\cite{Ly2004,Ly2007,Wa2008} and confirmed in VLBA 15 GHz
images \cite{Ko2007}.  These structures are also seen at
86~GHz~\cite{Hada2016}.  All of the high frequency VLBI observations of
M87 show that the edge-brightened structure has a parabolic shape
described by $r \propto z^{0.58}$ \cite{SN2012}.  That parabolic shape
continues to a de-projected distance of about $10^5$ R$_s$,
(\cite{SN2012} assumed an angle to the line-of-sight $\theta =
14\degree$) near the structure named ``HST1'', where it changes to
conical.  Unlike the structure with a radio core significantly offset
from the black hole that is thought to be the situation in many
blazars \cite{Mar2008}, the radio core in this much weaker source
appears to align with the black hole to within roughly 20 R$_s$
(deprojected with $\theta \sim 20\degree$) as evidenced by the
counterjet structure and by astrometric evidence for the expected core
shift with frequency induced by the optical depth \cite{HadaCore2011}.

The speed of the jet has been reported as subluminal on small scales
\cite{Ko2007,Hada2016} and as mildly superluminal ($\sim$2c)
\cite{Wa2008}.  A region of acceleration was reported on scales of a
fraction of an arcsecond~\cite{Asada2014}, well beyond the region
covered by the 43 GHz observations. The highest superluminal speeds
seen in M87 are about 6c as seen at HST1 in optical \cite{Biretta1999}
and radio \cite{Cheung2007} observations.  The~presence of a
counterjet that quickly drops below detectability is additional
evidence for relativistic speeds beyond the inner couple of
milli-arcseconds (mas).

In this contribution, we focus on measurement of the speed of the jet
over the inner several mas based on the 2007 and 2008 observations and
on results from the roughly annual observations between 1999 and 2016.
The annual observations demonstrate side-to-side motions of the whole
jet with a time scale of roughly 9 years.  More extensive
presentations of the results from the 43~GHz VLBA project will be
given elsewhere soon. The main data paper with some
discussion will be Walker \mbox{et al. \cite{Wa2016}}.  The wavelet-based
velocity measurements, and much interpretation and model fitting, are in \mbox{Mertens et al. }\cite{MLWH2016}.

\section{The Observations}

The 43 GHz VLBA M87 project broadly encompasses observations made
under several proposals between 1999 and the present.  The sensitivity
of the array increased significantly during this period mainly due to
bandwidth increases, especially the factor of 4 improvement in maximum
bandwidth and factor of 8 improvement achieved for this project before
the 2013 observations.  The primary goal of the project was an
explicit attempt to measure the jet speed by shortening the interval
between observations.  A pilot project to determine the best interval
was conducted in 2006.  Throughout 2007, M87 was observed every 3 weeks.
Despite the pilot, that interval was determined to undersample the
motions, so observations were made about every 5 days for 2.5 months
in 2008.  Unfortunately, the 2008 data quality is lower than for the
2007 data because of limitations in the ability of dynamic scheduling
to avoid poor observing conditions when the observing dates are
tightly constrained.  Thus the 2007 movie, despite its limitations,
remains the best resource with which to study the motions in the
source \cite{Wa2008}.

The 2008 data fortuitously corresponded to a significant rise in flux
density (by about 74\%) from the unresolved core region---the
largest flare actually caught in recent years.  That flare coincided
with a flare in the TeV energy regime that was observed by VERITAS,
H.E.S.S.  and MAGIC \cite{Acciari2009}, strongly suggesting that the
TeV emission was produced in the same location as the radio emission
very close to the black hole.  Since then, an on-going project to
catch other correlated radio-TeV flares has resulted in roughly annual
VLBA observations, made to check the source status near the start of
each TeV observing season.  There were also multiple sessions in 2010
and 2016 as a result of high energy triggers.  Significant flaring was
not seen, but multiple epoch images of the source structure were
obtained and will be reported in Walker et al. \cite{Wa2016}.  In
2012, a TeV flare that did not reach the trigger level for the 43 GHz
project was followed up at lower frequencies and a radio flare has
been~reported \cite{HadaFlare2014}.

All of the data used for this project were collected by the VLBA
without the use of additional antennas.  The correlation was done on
the VLBA correlator in Socorro, NM.  Data reduction was done using
AIPS.  Because a side-goal of the project involved phase referencing
between M84 and M87 to obtain their relative proper motions
(successful, but to be reported elsewhere), the calibration included
several steps not strictly needed for M87 imaging based on
self-calibration.  Corrections were made for improved Earth
Orientation Parameters (EOP), for the ionosphere using models from the
geodetic community, for atmospheric delays using the AIPS task DELZN
(rates for older data, delays for data since 2008 when geodetic
segments started being added), for atmospheric absorption using sec(Z)
fits to the system temperatures in task APCAL, and for the bandpass
shapes.  The a priori gains provided by the VLBA staff were used for
amplitude calibration.  The imaging involved many iterations of self
calibration and CLEAN.  Some key capabilities of the AIPS task IMAGR
that enabled production of images without excessive CLEAN artifacts
were the robust weighting scheme and multi-resolution~CLEAN.

\section{Jet Shape}

\begin{figure}[H]
\centering
\includegraphics[width=13.5cm]{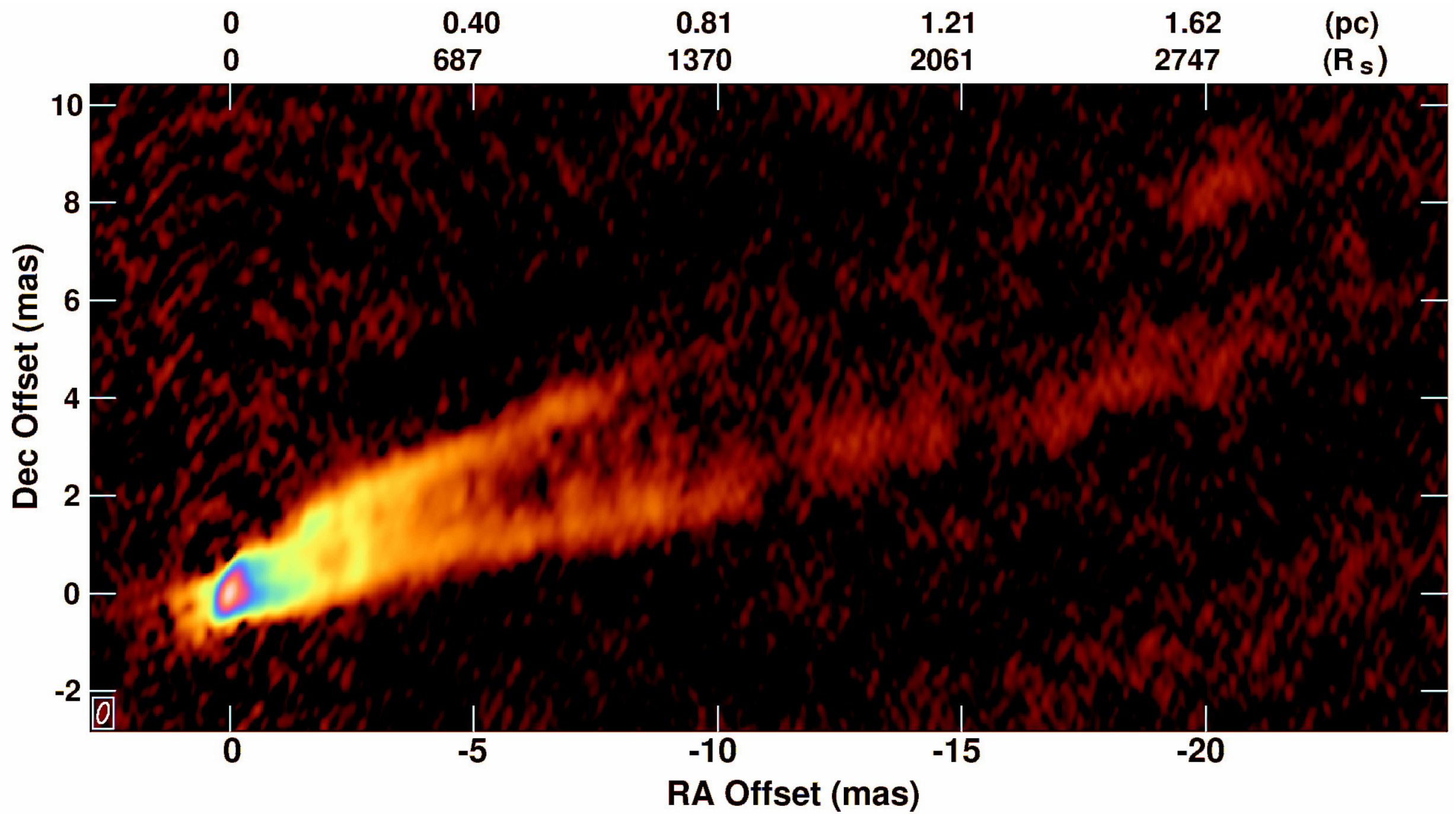}
\caption{An image of M87 made with the Very
Long Baseline Array (VLBA) at a frequency of 43 GHz
based on data taken on \mbox{12 January 2013} using the upgraded VLBA at a
recording bit rate of 2 Gbps.  The axes are in mas
from the brightest feature.  Above, the axis ticks are labeled in parsecs
and in Schwarzschild radii.  The~convolving beam is $0.43 \times 0.21$ mas
elongated along position angle $-16\degree$.}
\label{98Afull}
\end{figure}

To show the overall structure as seen by the VLBA at 43 GHz, one of
the best single-epoch images from the ongoing 43 GHz VLBA project
\cite{Wa2016} is shown in Figure~\ref{98Afull}.  This image benefited
from the use of the new wide bandwidth system on the VLBA so it has a
sensitivity comparable to the stacked image shown in other
publications \cite{Wa2008} but is not subject to the smearing inherent
in the stack.  The image clearly shows the edge-brightened structure
and presence of the counterjet that have been noted in other
publications as noted in the Introduction.  It also shows that the
source has fine structure that varies with time as demonstrated by the
smoother appearance of the stacked images.

\begin{figure}[H]
\centering
\includegraphics[width=12cm]{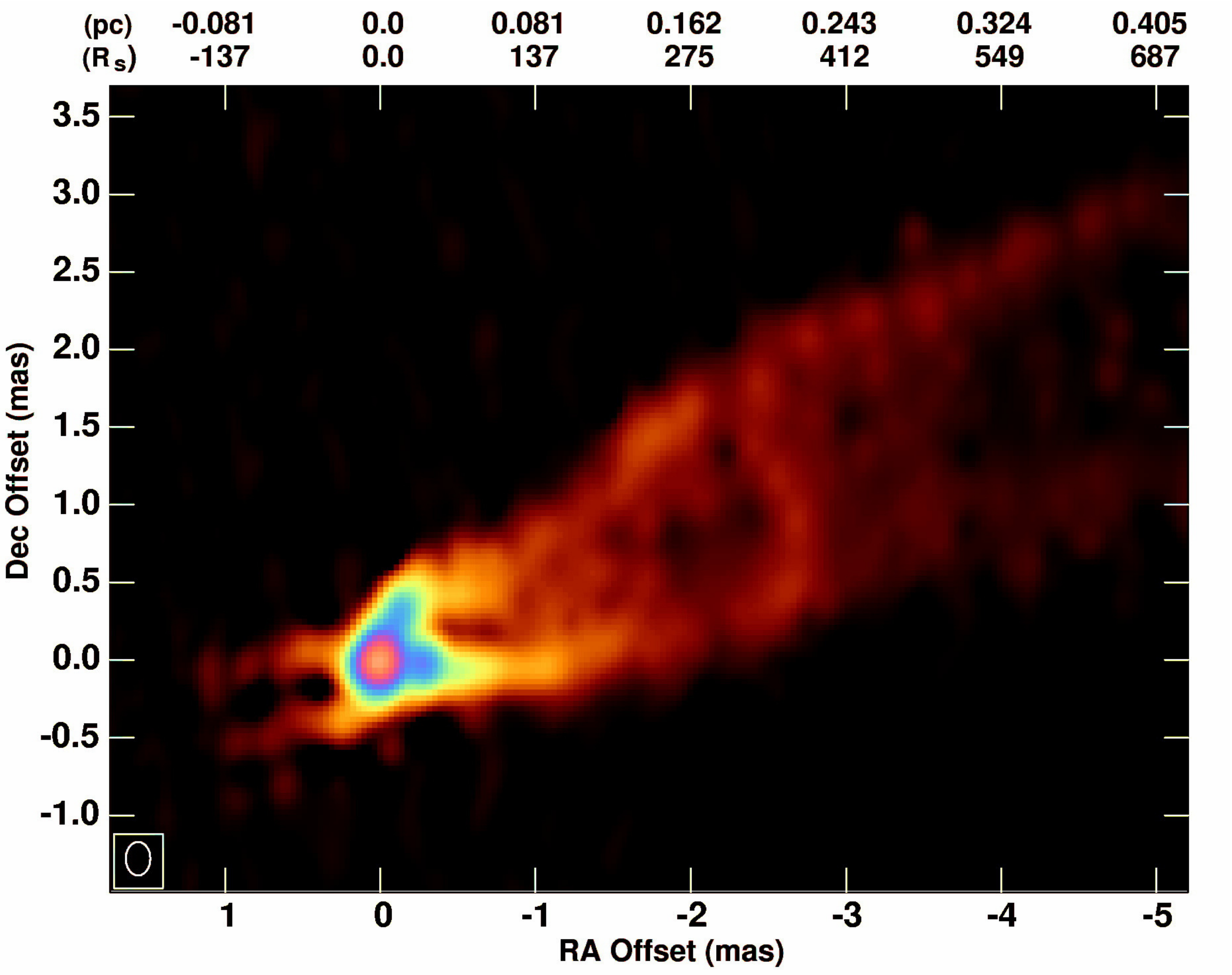}
\caption{An image of the inner portions of M87 based on the same 12 January 2013
 data used for Figure~\ref{98Afull}, but made with uniform
weighting to maximize resolution, and then superresolved by 30\% in
the N-S direction (convolved with a Gaussian beam that is 30\% smaller
than the central peak of the point spread function).  The convolving beam
used is $0.158 \times 0.215$ mas elongated along position angle
$0\degree$.  The axes are mas from the core.  The
higher resolution helps delineate the structure close to the core, and
emphasizes the symmetry between the jet and counterjet.  Note that the
sharp kink in the northern ridge at about 0.6 mas from the core is not
a lasting feature of the jet.}
\label{98Asup}
\end{figure}

To show the counterjet in more detail, a higher resolution version of
the same image, made with uniform weighting and 30\% superresolution
in the N-S direction, is shown in Figure~\ref{98Asup}.  It shows the
counterjet clearly and the symmetry of the jet and counterjet
structures.

\vspace{-6pt}
Analysis of the detailed structure will be given in upcoming publications
\cite{Wa2016,MLWH2016} while here we focus on the jet speed and long
term evolution.

\section{Jet Speed}

A rapidly changing structure for M87 is apparent in all of the VLBA 43
GHz data.  A~movie made from the first 11 epochs of the 2007
data, shown in the on-line material for \cite{Wa2008}, gives a strong
visual impression of rapid motions with an apparent speed in
projection of about 2c.  A~formal measurement of the speed was not
made at that time because the internal structure of the jet evolves
rapidly, making it somewhat difficult to identify clear components to
follow from epoch to epoch.  This~difficulty is compounded by the fact
that the 3-week intervals undersampled the motions.  Recently, two
methods were used to obtain better information on the velocity field
in the inner M87 jet.

The first method is a traditional effort to measure component motions.
For each of the 23 epochs from 2007 and 2008,
 the positions and peak
flux densities of the many emission peaks were measured.  The
positions were determined visually rather than by using formal fits
which are difficult due to the complex source and blending of
features.  An effort was then made, also visually, to relate peaks
between epochs.  By doing it visually, it was possible to take into
account adjoining structure such as pits and multiple peak features in
component identification.  For a significant fraction of the peaks, 
an~identification with peaks in other epochs was not clear so they are
not included in the velocity analysis.  We caution readers that this
method does suffer from the possibility that observer bias will affect
the outcome.  Variations on the method, that automatically determine
the related features, give~similar results.

\begin{figure}[H]
\centering
\includegraphics[width=12cm]{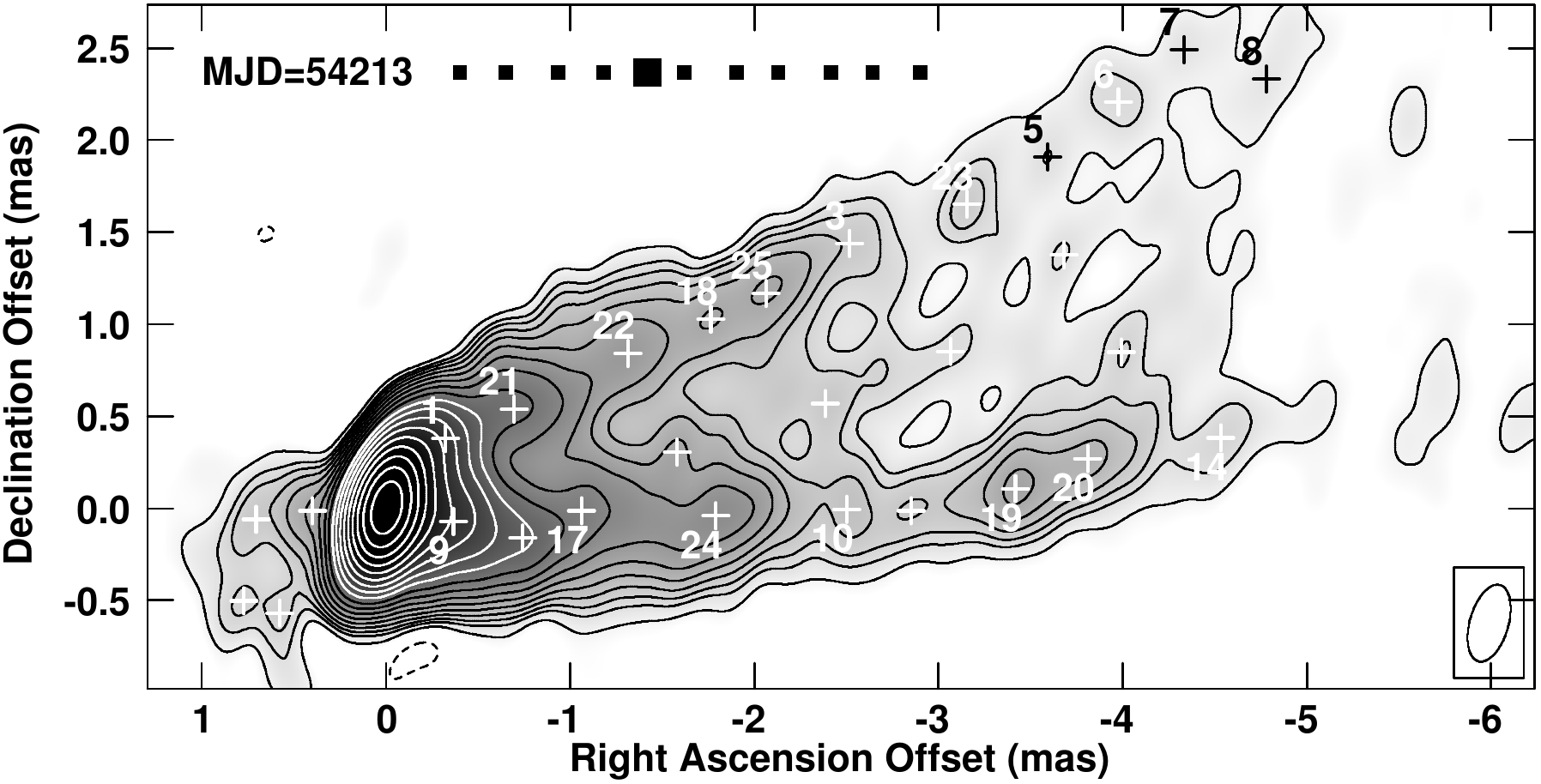}
\caption{An example VLBA image of M87 at 43 GHz.  This one is based on
data taken on 23 April 2007.  The lowest contours are at --1, 1, 2,
2.83, 4 mJy/beam increasing from there by factors of $\sqrt{2}$.  The
convolving beam is $0.43 \times 0.21$ mas elongated along position
angle $-16\degree$.  The crosses mark positions of local maxima
identified visually.  Images like this were created, and examined for
local maxima, for 23 epochs in 2007 and 2008.  The numbers identify
maxima that visual examination suggests can be identified at multiple,
sequenctial epochs.  All of the images will be shown in Walker
et. al. \cite{Wa2016}.}

\label{2007apr}
\end{figure}

\begin{figure}[H]
\centering
\includegraphics[width=12cm]{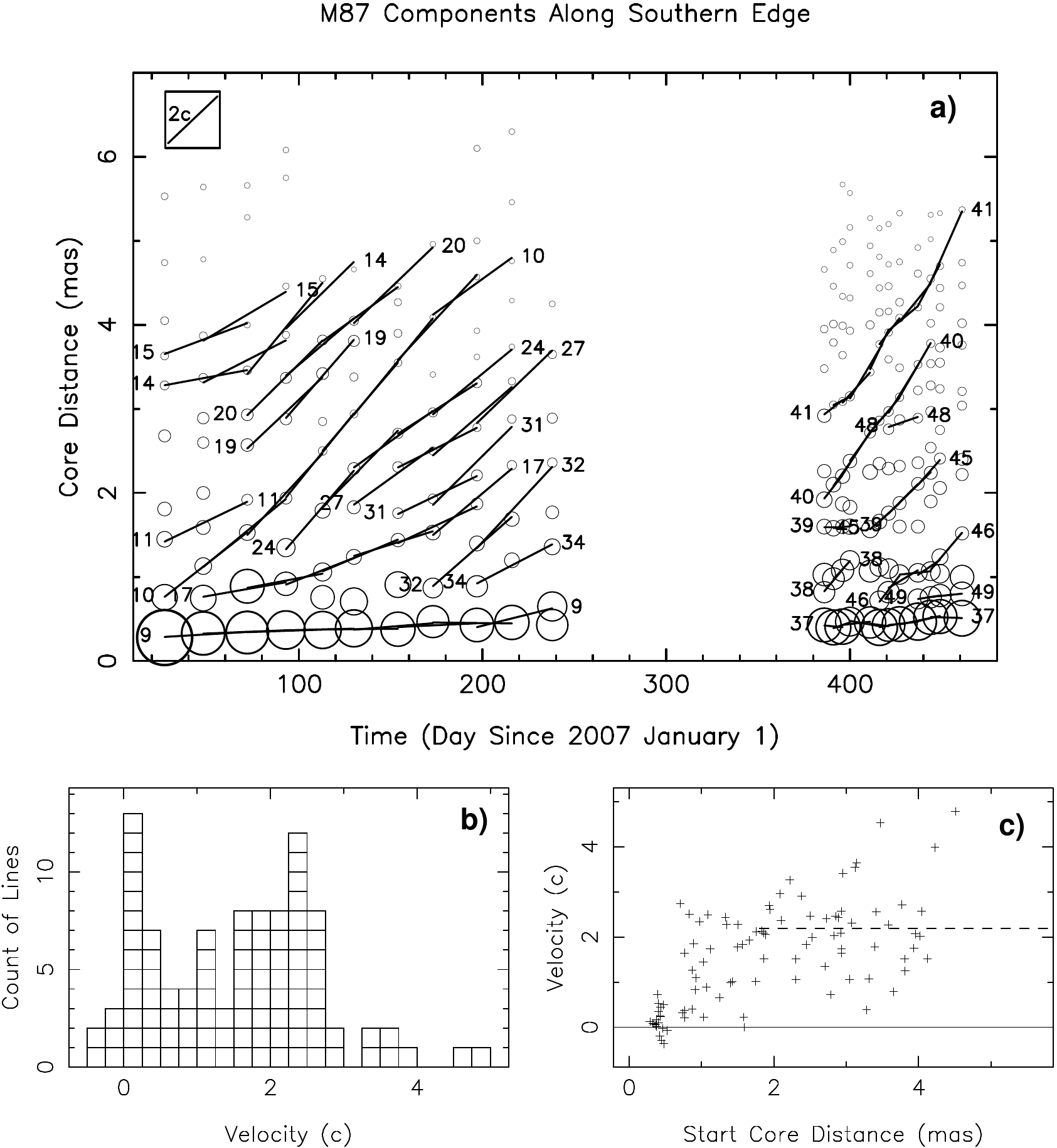}
\caption{The data, based on visual determination of component
positions and of associations of components between epochs, that was
used to characterize the component speeds along the southern edge of
the jet.  Similar data for the northern edge will be given in
reference \cite{Wa2016}.  (\textbf{a}) The symbols show the positions of the
peaks such as those shown in Figure~\ref{2007apr} with the symbol
sizes (areas) proportional to peak flux density.  The numbers indicate
components that were visually identified to correspond from epoch to
epoch.  The lines represent fits for position and speed for groups of
three peaks for the presumed same feature from adjoining epochs;  (\textbf{b}) A
histogram of speeds from the three-peak fits;  (\textbf{c})~The speeds from the
three-peak fits as a function of distance from the core of the first
peak.
}
\label{Speed}
\end{figure}

To determine speeds, a least squares fit was done to each set of
related features from three adjoining epochs.  The statistics of the
measured speeds were then examined for trends.  This was done
separately for the north and south sides of the jet.  There were some
peaks in the middle region of the jet and in the counterjet, but not
enough for a reasonable analysis of speeds.  A sample image, with the
visually identified peaks marked, is shown in Figure~{\ref{2007apr}}.
The core separation of all the peaks on the southern side, with a
connecting line representing the fit to each set of three related peaks,
is shown in Figure~{\ref{Speed}a}.  Figure~{\ref{Speed}b} is a
histogram of the measured speeds of the 3-peak segments while
Figure~{\ref{Speed}c} is a plot of those speeds as a function of
the core distance at the start of the line segment.  Both
Figure~{\ref{Speed}b,c}  show that there is a range
of speeds, with emphasis on nearly stationary peaks and peaks moving
at an apparent speed of somewhat over 2c.  It appears that peaks
accelerate over the inner 2 mas from very slow speed near the core to
higher speeds further out.  The acceleration is consistent with the
trend of the jet/counterjet sidedness ratio and explains why the
counterjet is apparant close to the core, but rapidly fades to below
detectability as the relativistic beaming increases.

\begin{figure}[H]
\centering
\includegraphics[width=10cm]{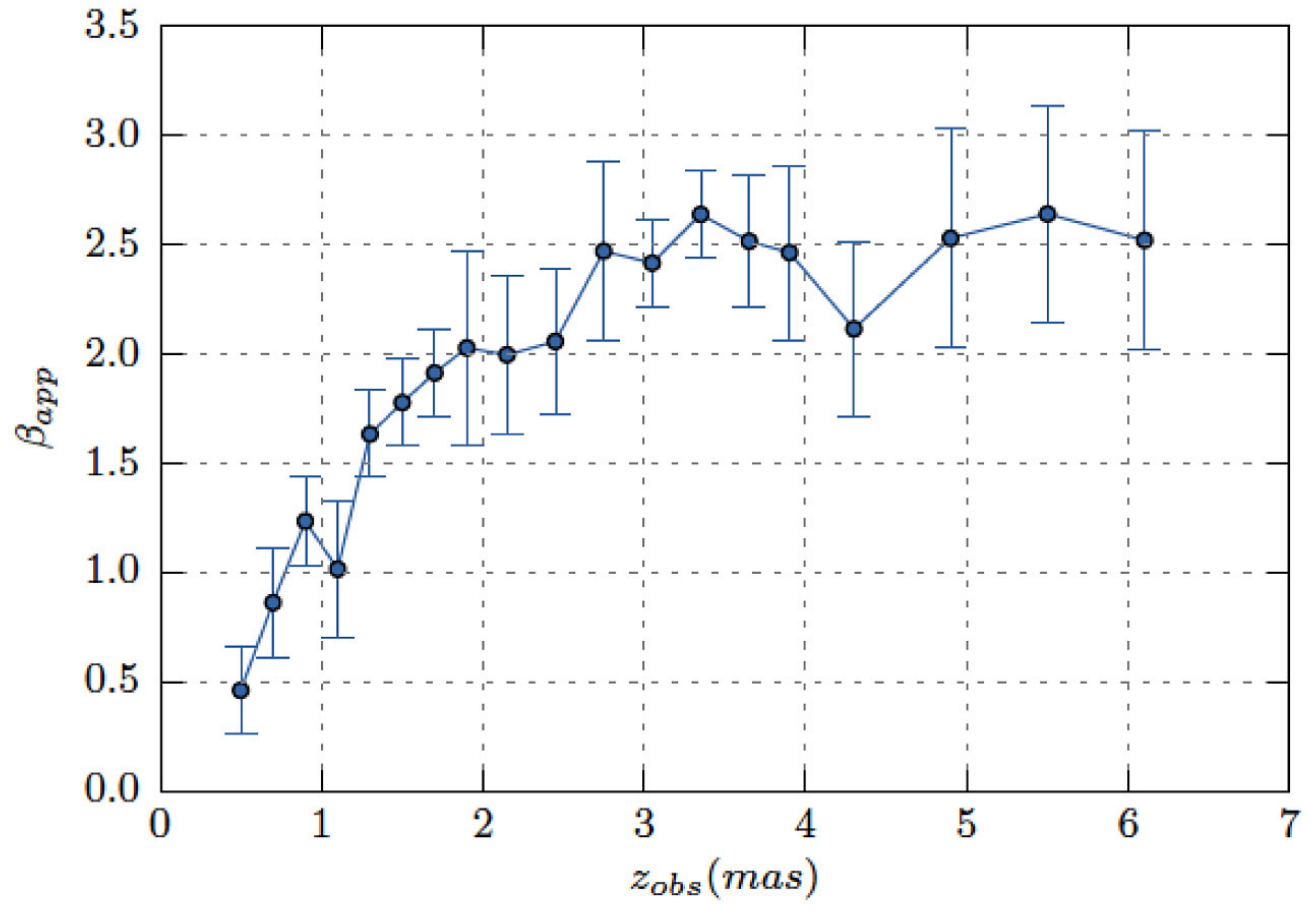}
\caption{The speed of the faster system found using the Wavelet Image
Segmentation and Evaluation (WISE) analysis
as a function of distance from the core \cite{MLWH2016}.
}
\label{WISEc}
\end{figure}

The second method used to obtain the jet speed uses the Wavelet Image
Segmentation and Evaluation (WISE) analysis \cite{ML2015}.  This work,
with significant analysis results, is presented in Mertens et
al. \cite{MLWH2016}.  The method is able to determine the velocity
field in considerable detail, including being able to detect multiple,
overlapping velocity fields.  A presentation of the WISE results on
M87 is given elsewhere in these proceedings \cite{L2016}.  In brief,
two velocity systems are identified.  One has a speed of $\sim$0.4c and
could be associated with an instability pattern or an
outer wind.  The other has a speed, at larger core distances, of
$\sim$2.3c with strong evidence for acceleration in the
inner 2--3 mas. The overlapping systems suggest either
stratification of the jet or the combined presence of features
following the bulk speed and features indicating patterns in the flow
such as shocks, instabilities, or~external influences.  A plot of of
the velocity as a function of core distance of the faster component,
that may be tied to the bulk speed, is shown in Figure~{\ref{WISEc}}.
The acceleration shown is consistent with the results shown in
Figure~{\ref{Speed}c}.

Other analysis results were obtained from the WISE velocity
field \cite{MLWH2016} and will only be mentioned briefly here.  The
speeds show a small difference for the north and south rims of the
jet.  This can be interpreted as a rotation with a rate of $\Omega
\sim10^{-6}~$yr$^{-1}$.  Such a rotation rate is consistent with the
launch of the observed portion of the jet from a region of the disk
about 5 R$_s$ from the black hole center.  MHD modeling of the
acceleration and collimation provides a good fit to the data for a
Poynting flux dominated case with equipartition between Poynting and
kinetic flux reached at about 3000 R$_s$.  Three~methods, the
sidedness and counterjet speed, the rotation analysis, and the MHD
fits, independently give an angle to the line of sight of $\theta
\approx 18\degree$.

\section{Long Term Variations}

\begin{figure}[H] \centering
\includegraphics[width=10cm]{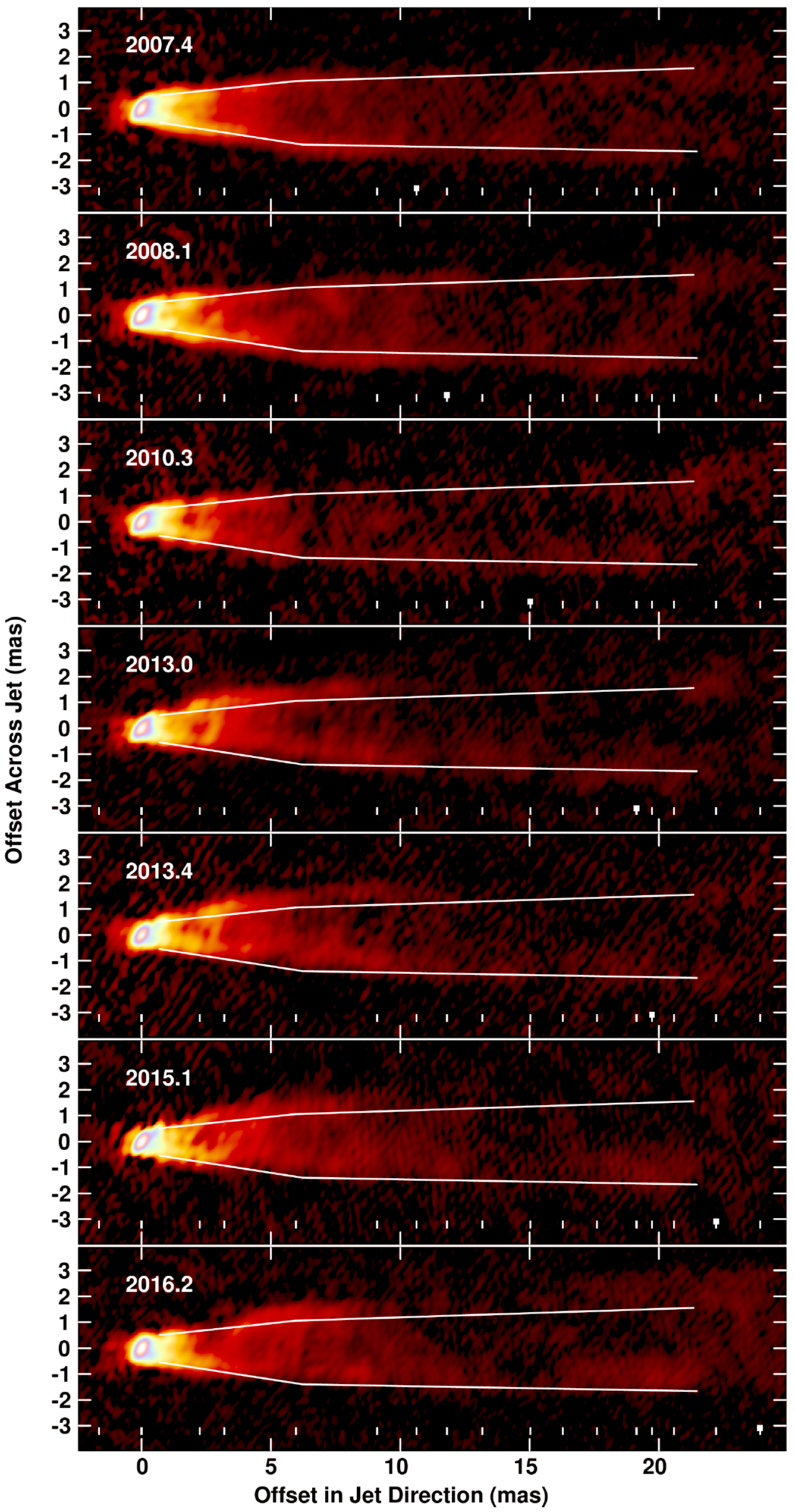} 
\caption{A montage of the seven best images or image stacks from the
 set of roughly annual VLBA 43 GHz images of M87 made between 1999 and
 2016.  The convolving beam is the same as for Figure~\ref{98Afull}
 and the images have been rotated in position angle by $-18\degree$.
 The lines overlaying the images mark approximately the ridge line of
 the bright edges of the jet in 2007.  The ticks at the bottom of each
 frame show where the frame (bold tick) is in the full sequence.
 Those ticks are especially useful in the movie available at the URL
 listed at the end of the summary and eventually in Walker et
 al. \cite{Wa2016}.}

\label{LongTerm} \end{figure}

The VLBA 43 GHz data sets span the period from 1999 to 2016 with
increasing image quality over time.  At least one image is available
for most of the years.  The images allow changes over
years to be detected.  Figure~{\ref{LongTerm}} shows a selection of 7
of the best images or image stacks from that period.  The~full set will be given in Walker et al. \cite{Wa2016}.  The first
3 of these images are stacks (noise-weighted mean) of multiple images
from data taken near the marked time.  The number of images stacked
are 11 for 2007.4, 12 for 1008.1, and 6 for 1010.3.  The other four
images are from single epoch data but benefit from the significantly
higher sensitivity of the upgraded bandwidth on the VLBA.

\begin{figure}[H]
\centering
\includegraphics[width=10cm]{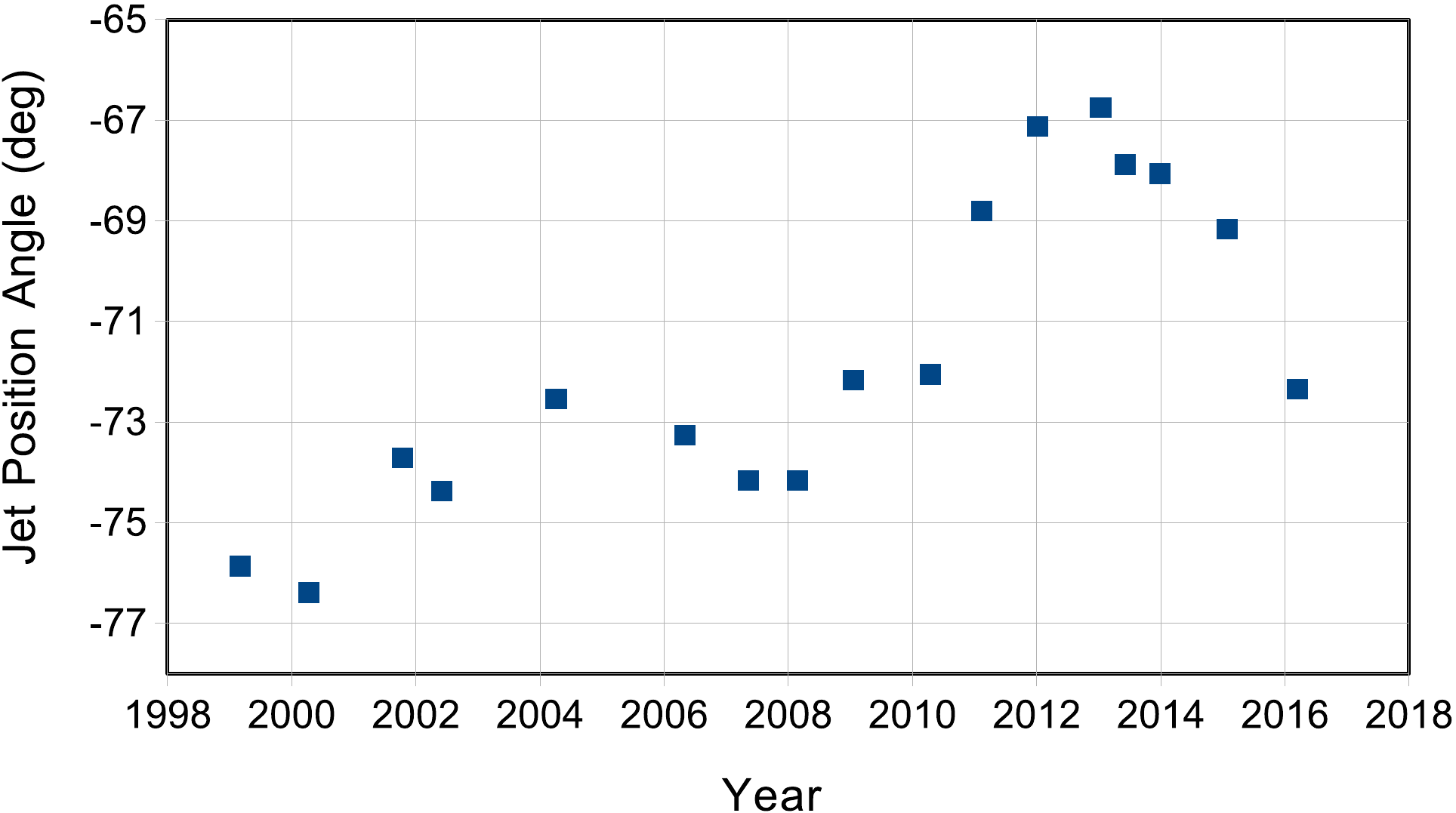}
\caption{The time history of the position angle (counterclockwise from
north) of a line between the core and the transverse midpoint of the
jet at a core distance of 3 mas.  The angle is measured directly from the
images, which show the jet in projection.  The points are suggestive
of a 9-year cycle superimposed on a drift.
}
\label{JetAngle}
\end{figure}

Sideways motions are seen in the annual images.  Lines have been
overlayed on each image in Figure~{\ref{LongTerm}} that approximate
the north and south ridges in 2007.  They are meant to make clear the
sideways translation of the jet observed in the later epochs.  Clearly
the images in 2013 and 2015 have the whole body of the jet (both
ridges) shifted to the north.  The shift north first appears in 2011
(in the full sequence) in the region about 2 to 3 mas from the core.
It then propagates outward at an apparent rate that has not been been
measured carefully yet, but appears to be $\sim$5 mas yr$^{-1}$ or
near 1.3c.  This is more than half the speed seen in the individual
components, suggesting the effect is nearly, but not quite, ballistic,
perhaps indicating a change in orientation of the jet launch.
Figure~\ref{JetAngle} shows the position angle, relative to the core,
of the transverse center of the jet at a core distance of 3 mas.  The
figure suggests an oscillation with a time scale of about 9 years
superimposed on a global drift, or a much longer oscillation.  With
less than two full periods, it is too early to tell if the position
angle variations are truly periodic, which would suggest precession.
An alternative could be quasi-periodic variations in jet direction as
has been seen in some 3D GRMHD simulations \cite{Tch2012}.

\section{Summary}

A number of conclusions about the M87 jet have been reached based on
the 43 GHz VLBA observations.  Some of these results have also
been reported by other investigators as cited in the discussions
above.  The main conclusions are:

\begin{itemize}[leftmargin=*, labelsep=5mm]

\item The jet has an edge-brightened structure with a wide opening angle
at the base and a parabolic shape over the region of these observations.

\item The velocity structure is complex, with co-existing slow
moving features ($\sim$0.4c) and superluminal features
($\sim$2.3c).  These are seen both using traditional methods
of following emission peaks and through the WISE analysis.

\item The faster moving system accelerates to near its full speed over
the first $\sim$2~mas, which is $\sim$0.16~pc ($\sim$274 R$_s$) in
projection or $\sim$0.52 pc ($\sim$890~R$_s$) along the jet for a
jet angle to the line-of-sight of $18\degree$.

\item There is a counterjet that mirrors the structure of the main jet,
but drops rapidly in brightness as would be expected in the acceleration
region.

\item An angle to the line-of-sight of $18\degree$ is determined by 3
methods based on the WISE results.

\item There is a difference in the measured speed of the northern and
southern rims of the jet.  A possible implication is rotation of the
jet with a rate of about $\Omega \sim 10^{-6}~$s$^{-1}$.  That, in turn,
suggests the jet is launched from the disk at a radius of $\sim$5~R$_s$.

\item Roughly annual observations over 17 years show that there are
side-to-side motions of the entire jet with a characteristic time scale
of about 9 years.

\end{itemize}

The WISE analysis is described in detail in Mertens et al.
\cite{MLWH2016}.  The main presentation of the images from the M87 43
GHz VLBA project will be in Walker et al. \cite{Wa2016}.  Both of
these works discuss the inferences that can be drawn from the data
concerning the nature of the jet in much more detail than has been
presented here.  Much of the data and the movies can be found at
\myurl{http://www.aoc.nrao.edu/~cwalker/M87/index.html}.


\acknowledgments{The Very Long Baseline Array is an instrument of the
National Radio Astronomy Observatory, which is a facility of the
National Science Foundation operated under cooperative agreement by
Associated Universities, Inc.  This work made use of the Swinburne
University of Technology software correlator \cite{DiFX}, developed as
part of the Australian Major National Research Facilities Programme
and operated under licence.  Florent Mertens was supported for this
research through a stipend from the International Max Planck Research
School (IMPRS) for Astronomy and Astrophysics at the Universities of
Bonn and Cologne.  Chun Ly is supported by an appointment to the NASA
Postdoctoral Program at the Goddard Space Flight Center, administered
by Oak Ridge Associated Universities and Universities Space Research
Association through contracts with NASA.}

\authorcontributions{
Walker, Hardee, Davies, Ly, and Junor are the long-standing group
doing the 43 GHz VLBA observations of M87.  Major contributions are
Walker: Planning, observing, processing, and publication.  Hardee:
Theory, interpretation, and writing the main paper.  Davies: Data
processing and M84-M87 proper motion.  Ly: Reduction and writeup of
pre-2006 epochs.  Junor: Observion and publication of the earliest
epochs.  The WISE processing and results are from the thesis of
Mertens, supervised by Lobanov.}

\conflictofinterests{The authors declare no conflict of interest.}

%
%

\renewcommand\bibname{References} 

\end{document}